\documentclass[pre,aps,twocolumn,superscriptaddress,floatfix]{revtex4}%%%%where rstrans is the template name

\usepackage{bm}
\usepackage{graphicx}
\usepackage{amsmath}
\usepackage{amssymb}

\begin{document}

%%%% Article title to be placed here
\title{Induced and endogenous acoustic oscillations in granular faults}

\author{%%%% Author details
L. de Arcangelis$^{1}$, E. Lippiello$^{2}$, M. Pica Ciamarra$^{3,4}$ and A. 
Sarracino}

%%%%%%%%% Insert author address here
\address{Department of Engineering,
University of Campania "Luigi Vanvitelli",  81031 Aversa (CE), Italy\\
$^{2}$Department of Mathematics and Physics, University of Campania "Luigi 
Vanvitelli", 81100 Caserta, Italy\\
$^{3}$
Division of Physics and Applied Physics, School of Physics and Mathematical 
Sciences, Nanyang,
Technological University, 21 Nanyang Link, Singapore 637371, Singapore\\
$^{4}$
CNR-SPIN, Department of Physics, University "Federico II", Naples, Via Cintia, 
80126 Napoli, Italy \\
}

%%%% Subject entries to be placed here %%%%
%\subject{granular media, acoustic fluidization, friction}

%%%% Keyword entries to be placed here %%%%
%\keywords{dynamical weakening, viscosity reduction, earthquake triggering}

%%%% Insert corresponding author and its email address}
%\corres{L. de Arcangelis\\
%\email{lucilla.dearcangelis@unicampania.it}}

%%%% Abstract text to be placed here %%%%%%%%%%%%
\begin{abstract}
%Wide attention has been devoted recently to the frictional
%properties of random media. The dependence of the frictional properties of confined granular systems
%on the external perturbations is of particular interest, as related to 
%the unexpected weakness of some faults. Indeed, this weakness has been attributed to the emergence of acoustic waves
%that promote failure by reducing the confining pressure through a mechanism known as acoustic fluidization,
%also proposed to explain earthquake remote triggering. This mechanism is extensively discussed via the numerical
%investigation of a granular fault model. Indeed the stick-slip dynamics is affected only
%by perturbations at a characteristic frequency,
%leading to dynamical weakening as failure is approaching. Acoustic waves at the same frequency
%spontaneously emerge at the onset of failure in the absence of perturbations, supporting the relevance of
%acoustic fluidization in earthquake triggering. We also report on the fluidization properties of a granular material subject 
%to mechanical vibrations: we show that, on
%increasing the perturbation frequency, a re-entrant transition is observed,
%as the system first enters a fluidized state, and then returns
%to a frictional state.

The frictional properties of disordered systems are affected by
external perturbations.  These perturbations usually weaken the system
by reducing the macroscopic friction coefficient.  This friction
reduction is of particular interest in the case of disordered systems
composed of granular particles confined between two plates, as this is
a simple model of seismic fault. Indeed, in the geophysical context
frictional weakening could explain the unexpected weakness of some
faults, as well as earthquake remote triggering. In this manuscript we
review recent results concerning the response of confined granular
systems to external perturbations, considering the different
mechanisms by which the perturbation could weaken a system, the
relevance of the frictional reduction to earthquakes, as well as
discussing the intriguing scenario whereby the weakening is not
monotonic in the perturbation frequency, so that a re-entrant
transition is observed, as the system first enters a fluidized state
and then returns to a frictional state.

\end{abstract}
%%%%%%%%%%%%%%%%%%%%%%%%%%%

%%%%%%%%%% Insert the texts which can accomdate on firstpage in the tag "fmtext" 
%%%%%

\maketitle

\section{Introduction}

Seismic occurrence is an intermittent phenomenon which is mainly
controlled by macroscopic friction. 
The observation of earthquake triggering induced by relatively small changes in the stress \cite{hill93,kilb2000,gomberg2004,gomberg2005} suggests that friction weakens when faults are subject to external perturbations.
This phenomenon can be
attributed to the presence of crushed and ground-up rocks
produced during past sliding events, known as fault gouge.  
This can be treated as a granular material which can act either as a lubricant
or, via the formation of strong force chains, can inhibit the relative
slip between the fault walls.  The two different behaviours correspond
to the double nature of granular materials which can be found either
in a fluid like 'unjammed' state or in an amorphous solid 'jammed'
state. An earthquake can therefore be interpreted as an unjamming transition
from a jammed state, in which the gouge resists the existing stresses,
to a flowing one.  
The transition from jammed to unjammed
states plays a central role not only in earthquake triggering but also 
in the physics of avalanche
dynamics~\cite{herrmann} as well as in the manufacturing process in
material, alimentary and pharmaceutical industries~\cite{coussot}.
In the latter case understanding the role  of external perturbations takes on a great practical relevance, due to the
possibility of controlling the mechanisms which reduce friction or
enhance fluidization in different
conditions~\cite{persson,krim,rmpvanossi}. 

External perturbations are well known to be able to induce the transition from a jammed to a flowing state
in confined granular systems experiencing a high normal stress and a small shear stress, and which would
therefore be jammed in the absence of perturbations. This perturbation induced unjamming transition
might actually explain the shaking induced fault weakening ~\cite{UvH05,urb2004,JJ05,DB05,DB06,JSKGM08,BJM08,BJJ08,CVVZ09,krim,GLPC12,CRBUF11,GDGJMC11,DWDDH11,VEBBJ12,CVVZ12,GFGDJMC13,XHM13,maksephyschemestry,LELC15,ROLJJ15,LGV15,GSdALP15, corwin2015,LEC16,WDH16,dGW17}. 
Interestingly, external perturbations can fluidize granular systems through two distinct mechanisms.
On the one side, external perturbations in given amplitude and frequency ranges can produce the detachment between the confining plates and the fault gouge. When this occurs the gouge does not oppose to the movement of the plates, which therefore flow. 
 This behaviour can be understood, at least at the qualitative level,  ignoring internal degrees of freedom of individual grains and treating the granular material as a single block, i.e. describing the fault as a block subject to a normal and a shear stress,
 resting on a oscillating plate. This simplified description permits analytical results which  can be used to explain the non-monotonic behaviour of the friction as function of the vibration frequency \cite{CVVZ09,CVVZ12,GLPC12,prl2018}.

On the other side, external perturbations can also induce fault weakening in the absence of detachment.
In this case the single block description is no longer useful and  
it is necessary to take into account 
the reorganization of the contact force network among grains induced by sound propagation.  This reorganization can explain the experimentally observed modulus softening  \cite{JJ05} and hysteretic behaviour identified for sufficiently large acoustic perturbations \cite{BJM08,BJJ08}.  
Among several mechanisms proposed to explain fault weakening induced by transient waves, 
in this Chapter we focus on the hypothesis of Acoustic Fluidization (AF). The AF concept was introduced by Melosh in
1979 to explain the transition from simple to complex craters on the
Moon \cite{Mel79} and also the low coefficients of friction observed in large-volume rock avalanches \cite{Mel86}.
According to Melosh \cite{Mel79,Mel96} dynamic fault weakening can be attributed  to the  activation of short-wavelength vibrations in the fault core. 
%Indeed, for infinitely hard grains, when a
%network of forces is formed, the system is stuck unless external
%conditions are changed. Conversely, if grains are deformable the
%contact network exhibits vibrational modes. 
These  could generate stationary  oscillations at a characteristic frequency $\omega_{AF}$  counteracting the applied stress and eventually 
``lubricating'' the system. The AF mechanism is therefore expected to strongly depend on the value of the confining stress as well as on the thickness of the granular medium which affects $\omega_{AF}$.  
This scenario is consistent with laboratory experiments on confined grains which have documented  \cite{XHM13} a clear transition, induced by an external shaking,  from elastic like to fluid like behaviour. 
These experiments, indeed, find that the unjamming transition is dramatically controlled by the thickness of the sample and the striker impact velocity, which modifies the confining pressure, in agreement with the AF scenario. The AF mechanisms has been also proposed to explain the change in density of flowing materials \cite{VEBBJ12} as well as the velocity-weakening of macroscopic friction \cite{dGW17}.  

In this chapter we consider the different fludization mechanisms in numerical models of a granular fault. At variance with experiments, numerical simulations allow us to follow the trajectory of single grains obtaining insights not available in experiments. For example, numerical studies \cite{FGGJMC13} have shown that vibrations affect both weak and strong contacts among grains. However 
when the perturbation is  switched off the weak contacts essentially returns to their initial state while the strong ones remain weakened.
%which however can be regenerated as soon as the external perturbation is switched off, as well as strong contacts which conversely remain weakened compared to their initial state after removal of vibration. 
The weakening of strong contacts can be responsible of the clock advance of large slip events induced by the shaking. Furthermore, in numerical simulations it is also possible to explore the endogenous activation of oscillation at the frequency $\omega_{AF}$. This is the mechanism proposed by Melosh to justify the unexpectedly small ratio
between shear and normal stress in real fault systems.  
The idea is that the elastic strain accumulated inside the fault is able to excite short wave-lengths  which
in turn can promote the activation of additional vibrations inside the fault core leading to self-fluidization.  

In the following, we first
%Since results of numerical simulations represent the benchmark of this chapter we devote Sec.~\ref{model_fault} to 
present the details of the considered  model for the granular faults, and briefly discuss  some of its static features.
Then in Sec.~\ref{pert_freq} we focus on the conditions under which the external perturbation induces the detachment
between grains and the confining plate. In particular we use the results of a single block dynamics to explain the 
outcomes of experiments measuring the angular velocity $\omega$ of
a vane coupled to an external motor and embedded inside a confined
granular material. 
In Sec.~\ref{acoustic_flu} we focus on the AF mechanism, considering the role of
of external perturbations not able to detach the confining plates, 
and on the endogenous activation of acoustic vibrations at the onset of stick-slip
instabilities.  Conclusions are drawn in the last Section.

\section{A model for granular faults}
\label{model_fault}

Geophysical phenomena as earthquakes occur over very large length 
scales and involve high values of energy, implying that it is clearly 
impossible to reproduce them at the laboratory scale. Nevertheless, earthquakes are 
characterised by scale-invariant laws, as the ubiquitous Gutenberg-Richter 
law for the magnitude distribution, which indicates the absence of 
characteristic scales. This implies that the experimental and numerical 
investigation of model seismic faults might give access to the relevant 
physical properties of real faults. Many experiments and simulations of model 
seismic faults are therefore present in recent literature.

\begin{figure}[h!]
\includegraphics*[scale=0.3]{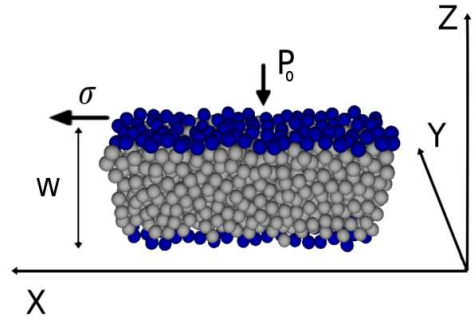}
\caption{
The fault gouge is enclosed between two rigid rough plates of dimension 
$L_x \times L_y$. Each plate is made of $L_x L_y/d^2$ spheres of
diameter $d$ placed in random positions in the $xy$ plane. Spheres are shifted
by a random $\delta z\in [0,d/2]$ in the perpendicular direction and then glued
to each other. In order to make the plates rigid, the particles keep their
relative positions. This preparation protocol ensures the roughness of both  confining plates. 
Molecular Dynamics  simulations are then performed for a certain time interval (thermalisation) 
during which if a particle of the granular bed has a very strong  impact with a rigid plate it remains glued on it. 
While the bottom plate is kept fixed, the top one is subject to a
constant pressure $P_{0}$ and attached to a spring of elastic constant $k_d$. 
The contact force model is described in Sec.\ref{num}.}
\label{figs1}
\end{figure}

\subsection{Numerical model}\label{num}
The numerical model of seismic fault presented here focuses on the fault core dynamics, where most of the 
shear displacement occurs. Results are obtained under the assumption 
that the region outside the core is a rigid body, so that the fault of width $W$ can be modelled as 
two parallel rigid plates of area $L_x L_y$ confined by a normal stress 
$P_0$ and subject to a shear stress $\sigma$ (fig.\ref{figs1}). See~\cite{Giacco2014b,3ddegriffa2014} for results
considering elastic confining boundaries.
The fault gouge, that in real faults consists 
of rocks produced in past wearing events, is modelled as a collection of 
frictional granular particles over a width $W$. Particles are mono-dispersed spheres of mass $m$ and 
diameter $d$. This choice does not lead to crystallization because of the rough
confining plates, however poly-dispersed particles would represent a more
realistic modelization of real gouges, numerically more demanding. 
In real faults the shear stress $\sigma$  slowly increases as a consequence of 
the convective motion of the upper mantle, and quickly decreases when a fault 
slips. In simulations, an analogous scenario is realised by applying the 
stress through a spring mechanism.
One end of the spring is attached to the plate, and thus moves with the 
plate velocity $v_p(t)$, while the other end is driven with a constant velocity 
$V$.
Assuming the spring to have zero equilibrium length and zero length at time $t 
= 0$, the force exerted by the spring at time $t$ is $F_t = k_d (Vt-x(t))$, 
where $x(t) = \int_0^t v_p(t') dt'$ and $k_d$ is the spring stiffness. Then 
the shear stress results to be $\sigma_{xz} = \sigma = F_t/L_x L_y$.
The stick-slip dynamics is recovered when the time over which the stress 
increases, which depends on the shear rate $k_d V/L_x L_y$, is much longer than the duration of a 
slip event, which depends on the pressure $P_0$ as well as on the dissipative 
properties of the system.
In numerical models, the confining plates are rough rigid objects, generally 
realised as a dense disordered assembly of granular particles. The 
plate is rigid since all relative distances between the particles in the 
plate are kept constant and particles in the plate are not allowed to rotate.

The interaction force between two granular particles has a normal and a 
tangential component. To describe the interaction, 
two particles $i$ and $j$ are considered, with radii $R_i$ and $R_j$, 
in position $r_i$ and $r_j$, linear velocities $v_i$ and 
$v_j$ and angular velocities $\omega_i$ and $\omega_j$. The particles interact 
only when in physical contact, i.e. when $|r_i-r_j| < d_{ij}=R_i+R_j$, 
i.e. when the overlap $\delta =d_{ij}-|r_i-r_j| > 0$.
We use the Harmonic spring-dashpot model~\cite{silbert},
\begin{equation}
{\bf F}_{ij} = {\bf F}_{ij}^{n} + {\bf F}_{ij}^{t} = \left[ k_n\delta {\bf n}_{ij} -m_{\rm eff} 
\gamma_n {\bf v}_{n})\right]   + \left[ - (k_t {\bf t}+m_{\rm eff} \gamma_t {\bf 
v}_{t}) \right].
\label{eq:interaction}
\end{equation}
The first term of Eq.~(\ref{eq:interaction}) corresponds to the normal interaction. 
Here $k_n$ is the stiffness of the particles, whereas $\gamma_n$ controls 
the dissipation and $m_{\rm eff}$ is the effective mass of spheres with mass $m_i$, 
$m_j$. In this model, the normal collision between two grains 
dissipates a constant fraction $(1-e^2)$ of the overall kinetic energy, where 
$e(k_n,\gamma_n,m_{eff})$ is the restitution coefficient. 
The second term of Eq.~(\ref{eq:interaction}) corresponds to the tangential 
interaction force, where $k_t$ is a tangential stiffness, ${\bf t}$ the tangential 
shear displacement, and $\gamma_t$ a damping parameter. The tangential shear 
displacement ${\bf t}$ is the integral of the relative tangential velocity ${\bf v}_{t_{ij}}$ 
at the point of contact, where 
${\bf v}_{t_{ij}} = {\bf v}_{ij} - {\bf v}_{n_{ij}} - 1/2( {\bf\omega}_i + {\bf\omega}_j) \times 
{\bf r}_{ij}$ depends on the angular velocities of the particles. Here the relative velocity is
${\bf v}_{ij}={\bf v}_{i}-{\bf v}_{j}$ and ${\bf r}_{ij}={\bf r}_{i}-{\bf r}_{j}$. The tangential 
force acting on a particle contributes to the overall torque acting on it, which 
induces its rotational motion.
The presence of a finite Coulomb's friction coefficient $\mu$ is implemented by 
introducing an upper bound for the tangential force, which is 
capped at $\mu |F_n|$. This is done by appropriately rescaling ${\bf t}$, thus 
mimicking the slipping of the contact. 
We measure the mass in units of $m$, the lengths in units of $d$ and time in 
units of $\sqrt{m/k_{d}}$. Unless specified, results are obtained 
for a restitution coefficient $e = 0.8$, a coefficient of static friction $\mu = 0.2$, the confining pressure is $P_{0}= {k_d/d}$, $k_{n}= 
2\cdot 10^{3}\, k_d$, $V=0.01\, d\sqrt{m/k_{d}},\gamma_t =0$, the temporal integration 
step of the equations of motion is $5\cdot 10^{-3}\,\sqrt{m/k_{d}}$ and  
$W \simeq 10\,d$ as in Ref.~\cite{prl2010}.
The model reproduces the stick-slip dynamics with slipping event sizes distributed
according to the Gutenberg-Richter law characterising real seismic occurrence,
independently of model parameters. In particular, the dynamics consists of almost periodic
large events, called slips, and of  power-law distributed smaller events, 
called microslips (Fig.~\ref{figE1}) \cite{epl2011}.

\subsection{Micromechanics of failure}

Given that the numerical model gives access to all possible quantities
of interest, it is useful to investigate how does the
system fail. There are two possibilities, both of them based on the
consideration that, as the shear stress increases, the inter-particle
forces slightly change, since the system is in mechanical equilibrium at
all times. In the local scenario, failure occurs as one of the
inter-particle contacts reaches the Coulomb threshold and starts
slipping. This local failure then propagates leading to the
macroscopic failure of the sample.  In the global scenario,
conversely, the shear stress increase leads to a
deformation of the energy landscape of the system. Thus the energy
minimum, in which the system is trapped when in a jammed configuration, might
gradually evolve into a saddle. When the minimum becomes a saddle, the
system fails. This second scenario is certainly at work in the absence
of frictional forces, that are known not to influence the statistical
properties of the model \cite{epl2011}. Formally, in the absence of
friction this second scenario occurs when an eigenvalue of the
dynamical matrix of the system vanishes. It has been shown \cite{prl2010}
that also in the presence of frictional forces failure results from a
global instability of the system.  Indeed, the system failure time
is the first time at which, if the shear stress is kept constant, 
at least one contact reaches its Coulomb threshold.

In frictional systems one cannot explicitly show that failure is a global 
process investigating the dynamical matrix of the system,
due to the absence of a Hamiltonian. As an alternative, 
the global instability scenario suggests an analogy with a second order phase 
transition, where a minimum evolves into a saddle point. On approaching the 
failure point one therefore expects the system to become increasingly more 
sensitive to external perturbations.
It is therefore interesting to investigate the evolution of the response of the system to 
external perturbations. 
A perturbation is considered as a decrease of the confining pressure $P_0$ 
by $\alpha P_0$ for a short time $\delta t_{pert}=0.1$, with $\alpha\ll 1$ or small enough 
to probe the linear response regime.
In order to guarantee the separation of time scales between the external drive 
and the mechanical relaxation, 
the external drive of the spring is kept constant by fixing $V=0$ when 
performing this study.
As a consequence of the applied perturbation, grains move. 
The response of the system is monitored by studying the susceptibility
\begin{equation}\label{1.1}
\chi_{\alpha}(t)=\frac{1}{\alpha 
P_0}\lim\limits_{\tau\to\infty}\left[\frac{1}{N}\sum_{i}\left({\bf 
{r}_i^{\alpha}(t+\tau)-\bf{r}_i^{0}(t+\tau)}\right)^2 \right]^{1/2},
\end{equation}
where $\bf{r}_i^{\alpha}$ and $\bf{r}_i^{0}$ are the asymptotic particle 
positions in the perturbed and unperturbed system, respectively. Here the
$t$ dependence indicates solely the time at which the perturbation is applied 
since the susceptibility is evaluated asymptotically 
and is therefore a static quantity. In the unjammed phase the susceptibility is 
divergent since global position rearrangements occur, whereas in the jammed 
phase only limited regions give contribution to $\chi_{\alpha}$ which therefore 
provides a measure of the correlation length.

By monitoring the response of the system for different $\alpha$ close
to slips and microslips a very different behaviour is
detected \cite{prl2010}. Indeed, close to microslips the response of the system is
linear in $\alpha$, namely $\chi_{\alpha}$ increases proportionally to
$\alpha$ as the microslip time is approached and drops to zero at the
subsequent jamming time. Interestingly, the $\chi_{\alpha}$ value at
the onset of the microslip is proportional to the microslip size,
indicating that the information on the slip size is contained in the
global state of particle positions.  Conversely, as the large slip
occurrence time is approached, the susceptibility suddenly diverges,
the sooner the larger $\alpha$. Therefore, in the limit of vanishing
perturbation $\chi_{\alpha}$ diverges at the slip time.  This
divergence, reminiscent of the response of a system close to a
critical point, indicates that global rearrangements of the particle
positions occur leading to a correlation length of the order of the
system size.  In order to better simulate the mechanical conditions of
seismic faults, the system should undergo periodic perturbations
representing the passage of seismic waves triggered by earthquakes
just occurred even far in space.  The frequency dependence of the
response will be discussed in the next Sections.

\section{Oscillations inducing the detachment between grains and the confined plates.}
\label{pert_freq}

In this section we review some recent results on the response of granular
packings to mechanical perturbations focusing on the role played by
the vibration frequency on the frictional properties of the medium.
In particular, we first discuss a single block model under vibration, to introduce the issue of friction reduction
in a simple context.
Then we consider a granular system similar to the gouge fault model introduced in Sec.~\ref{model_fault}
analysing the central role played by the vibration frequency.
Finally we report some recent experimental results on shaken granular systems
in a different setup, suitable for rheological studies, where a vane coupled to an external motor, immersed in the medium, is used to probe its viscosity properties.    

\subsection{A case study: The single block}
\label{single_block}

\begin{figure}[!htb]
\includegraphics[width=3.5in,clip=true]{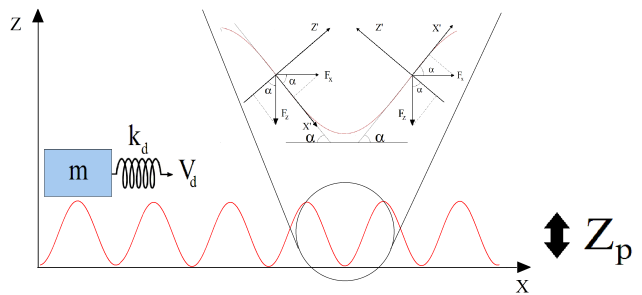}
\caption{Motion on a substrate with periodic corrugation. Magnification shows the frame of reference used to study the dynamic of the block. In this frame each force has a vertical and a horizontal component which depend on the angle $\alpha=\arctan\left(\frac{\partial{Z_{p}}}{\partial{x}}\right)$.}
\label{sist_primo}
\end{figure}

Friction between sliding objects is largely affected by mechanical
vibrations.  To introduce the problem on simple grounds, a numerical
study~\cite{GLPC12} on a spring-block model in the presence of
vertical vibrations is considered. This allows to bring to the fore some
fundamental ingredients that influence the frictional properties of
sliding solids, such as the external drive, the geometry of the
surfaces over which the block moves and the kind of confining force.
In particular, one observes non-trivial behaviour as a function of the
vibration frequency, such as friction reduction, that will be analysed
in more complex systems in Subsection~\ref{zap}.
The solid-on-solid model consists in a block of mass $m$ that is pulled
along the $x$ direction by a spring of elastic constant $k_d$ driven at constant velocity $V_d$ and moves
along a surface which is vibrated in the $z$ vertical direction according to
\begin{equation}
Z_p(x,t)=A \sin(2\pi f t)+A_x \sin(k_x x),
\label{zmaxp}
\end{equation}
where $Z_p$ is the vertical coordinate of the plate (see Fig. \ref{sist_primo}). One can
identify three main variants: In model A, the surface of the substrate
is flat ($A_x=0$) and the confining force is vertical; in model B, the surface
has a sinusoidal shape ($A_x>0$) and the confining force is still vertical; in
model C, the surface is sinusoidal ($A_x>0$) but the confining force is always
normal to it.
In particular, the equations of motion along the two directions $(x,z)$ with respect to a fixed reference system in model A are given by: 
\begin{equation}
m\ddot{z}=k_{n}(Z_{p}-z)\Theta(Z_{p}-z) - \gamma_{n}(\dot{z}-\dot{Z_p})\Theta(Z_{p}-z)-mg, 
\end{equation}
\begin{equation}
m\ddot{x}=-k_{d}(x-V_{\textrm{d}}t) - \gamma_{t}\dot{x}-k_{t}\mathcal{F}(\dot{x}, Z_p-z), 
\end{equation}
where $\Theta$ is the Heaviside step function, $k_{n}$ is the elastic constant, $\gamma_{n,t}$ are the  viscoelastic constants, normal and tangential respectively, while the quantity $k_{t}\mathcal{F}$ is a frictional term that is present when block and surface are in contact. It is assumed to be proportional to the shear displacement over the contact time interval $(t,t_0)$ between the block and the plate \cite{cundall} and is given by $\mathcal{F}(\dot{x}(t), Z_p-z) =\int_{t_{0}}^{t}\dot{x}(t')dt'$. The Coulomb friction is taken into account through the condition $|k_{t}\mathcal{F} |< \mu_s N$, where $N$ is the normal load; when this condition is violated $\mathcal{F}$ is set to zero. 

In model B and C we write the equations of motion in a frame of reference which moves along with the plate, with the horizontal axis tangential to the plate, see Fig. \ref{sist_primo}. In this frame the equations of motion for model B are 
\begin{eqnarray}
m\ddot{z}' &=& k_{n}(Z_{p}'-z')\Theta(Z_{p}'-z') - \gamma_{n}(\dot{z}'-\dot{Z_{p}'})\Theta(Z_{p}'-z')+\nonumber \\  &+& k_{d}(x'-V_{\textrm{d}}t)\sin(\alpha)-mg\cos(\alpha), \\
\label{zetaprimo}
m\ddot{x}'&=& -k_{d}(x'-V_{\textrm{d}}t)\cos(\alpha) - \gamma_{t}\dot{x}'-k_{t}\mathcal{F}(\dot{x}', Z_p'-z') \nonumber \\
&-&mg\sin(\alpha), 
\label{xprimo}
\end{eqnarray}
where $\alpha=\arctan\left(\frac{\partial{Z_p}}{\partial{x}}\right)$ at fixed time, while model C is described by
\begin{eqnarray}
m\ddot{z}' &=& k_{n}(Z_{p}'-z')\Theta(Z_{p}'-z') - \gamma_{n}(\dot{z}'-\dot{Z_{p}'})\Theta(Z_{p}'-z')+\nonumber \\  &+& k_{d}(x'-V_{\textrm{d}}t)\sin(\alpha)-P_l, \\
\label{zetaprimo2}
m\ddot{x}'&=& -k_{d}(x'-V_{\textrm{d}}t)\cos(\alpha) - \gamma_{t}\dot{x}'-k_{t}\mathcal{F}(\dot{x},Z_p'-z'),
\label{xprimo2}
\end{eqnarray}
 with $P_l$ a confining force which is always perpendicular to the surface.

\begin{figure}[!htb]
\includegraphics[width=2.6in,clip=true]{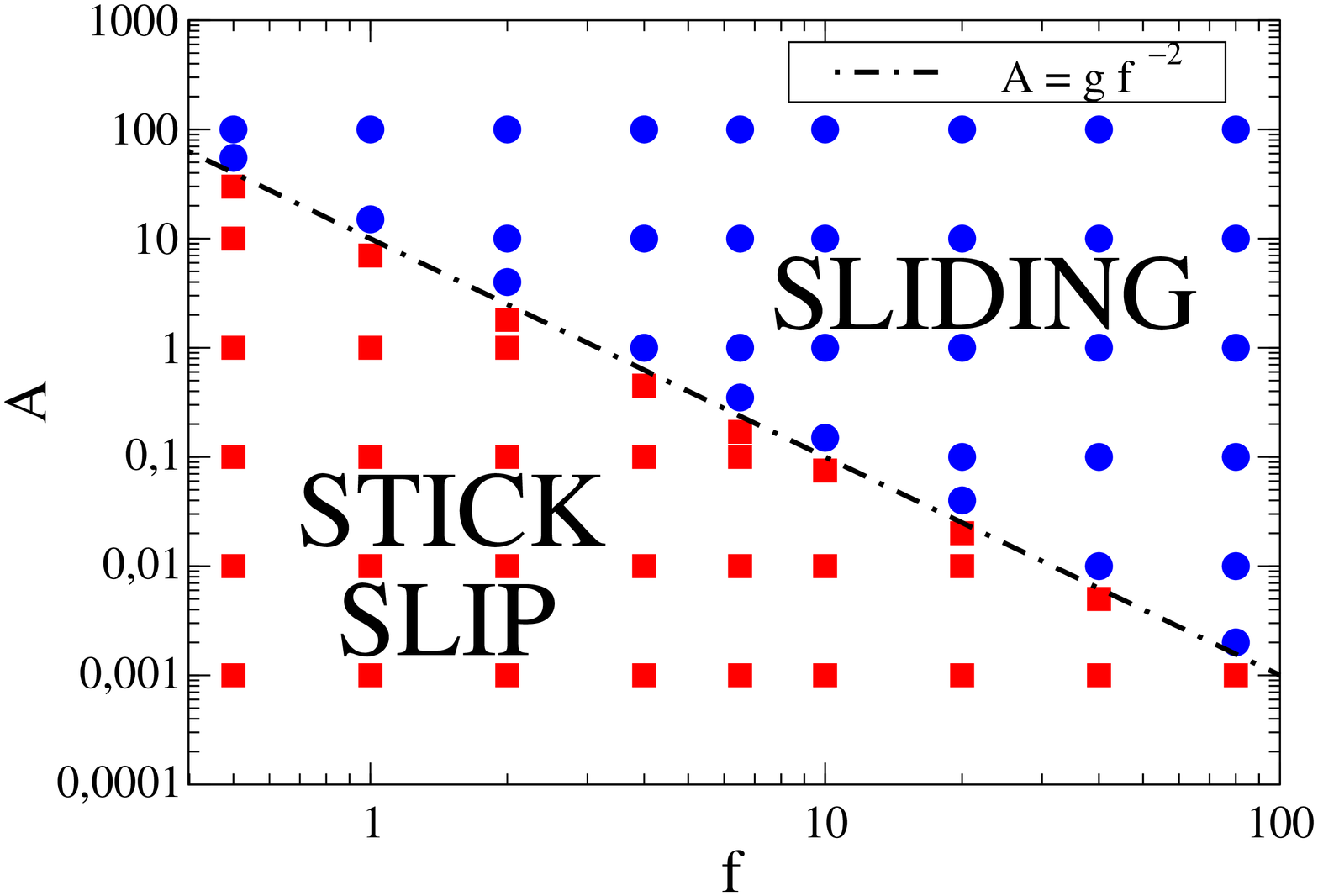}
\includegraphics[width=2.5in,clip=true]{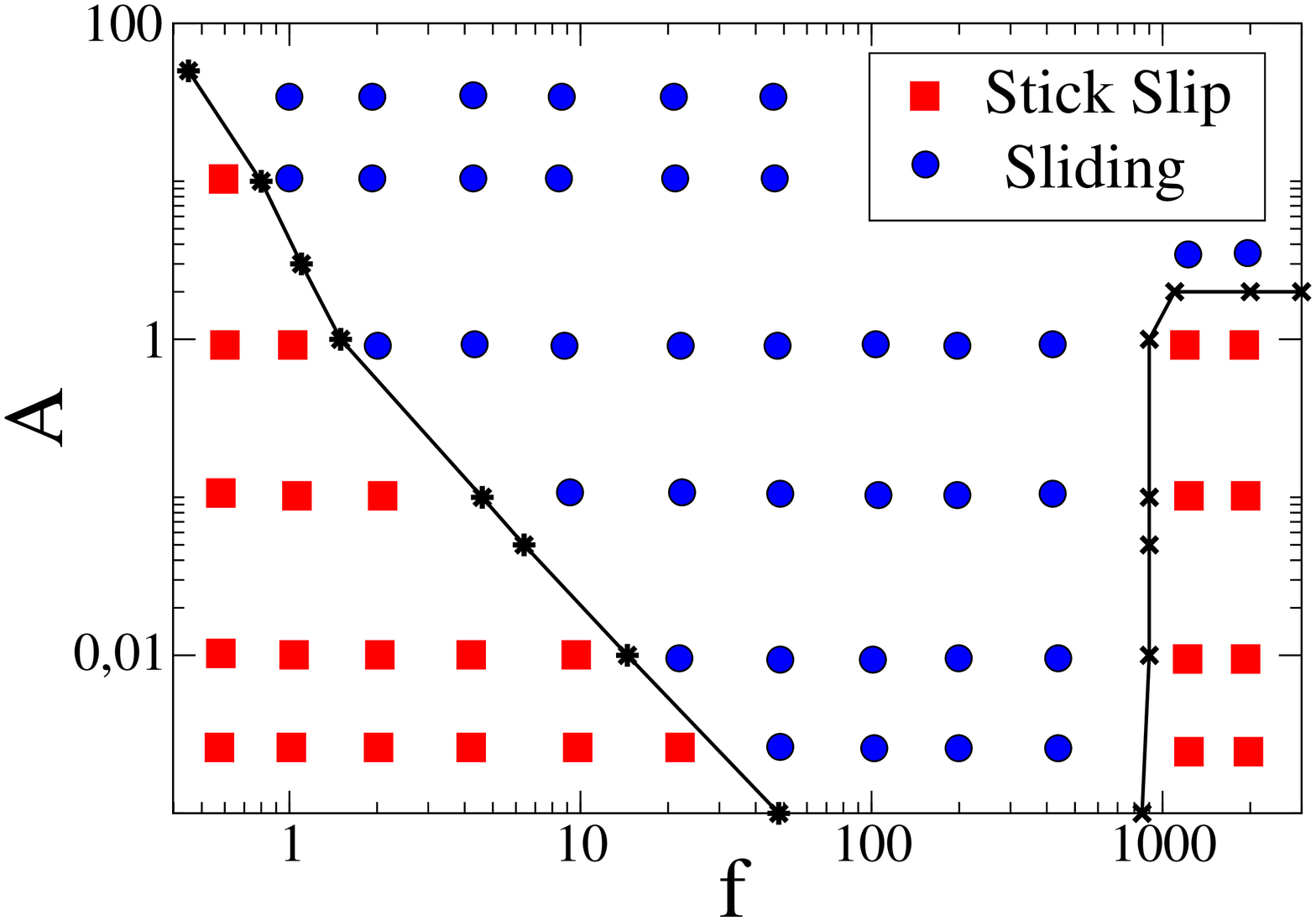}
\caption{Phase diagrams in the parameter space $(f,A)$ of the block
  models A (left) and C (right). Note the re-entrant transition at high
  frequencies in model C. }
\label{fig_block}
\end{figure}

Spring block models without vibrations present a sliding (fluid) phase
and a stick-slip (solid) phase~\cite{vasco} and the presence of
vibrations affects the transition between these two phases. The three
models introduced above show different behaviour: The main properties
of model B only depend on the driving velocity and are almost
independent of the presence of vibrations, whereas models A and C show
a transition from the stick-slip to the sliding phase for increasing
frequency (or amplitude) of the oscillating substrate, occurring when
its maximum acceleration overcomes the gravity acceleration $g$. Remarkably, in
model C one observes that a further increase of the frequency leads
to a second transition whereby the system re-enters the stick-slip
phase. More specifically, in order to better identify the transition one can introduce an order parameter, defined as
\begin{equation}
\phi=\frac{\langle(\dot{x}-V_d)^{2}\rangle_{t_a}}{V_d^2}, 
\label{xprimo}
\end{equation}
where the brackets $\langle\cdots\rangle_{t_a}$ indicate temporal averages over a period $t_a$. 
During the flowing phase, the block moves with the external drive velocity ($\dot{x}= V_d$) and therefore 
$\phi=0$. Conversely, in the stick-slip phase , $\phi$ takes a finite value that  
depends on the  period $t_a$ (or the number of occurred slips during $t_a$).  
The different behaviours of models A and C are summarized by the phase diagrams in the
parameter space $(f,A)$, obtained from numerical simulations in~\cite{GLPC12} and 
reported in Fig.~\ref{fig_block}. Model A is
characterized by the line $A=g (2\pi f)^{-2}$, separating a stick-slip region
and a sliding region that originates from the detachment
condition. The phase diagram of model C is more complex, featuring a
second transition from the stick-slip to the flowing phase at high
frequencies. This friction recovery transition is related to the
balance condition between inertial and dissipative
forces~\cite{GLPC12}. The results obtained for these single block
models indicate that friction recovery can occur even in simple
systems and can be related to the modulation of the surface over which
the block slides. In the following, a case is presented where
a similar phenomenology takes place in systems of granular particles.

\subsection{Complex rheological behaviour of a granular system under external perturbation}
\label{zap}

\begin{figure}[!htb]
\includegraphics[width=1.8in,clip=true]{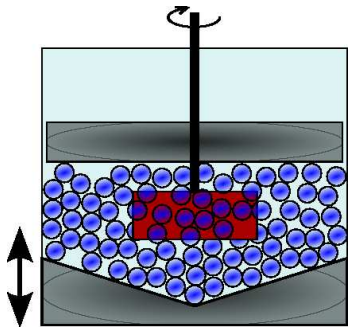}
\caption{Typical experimental setup for rheology. A driven vane (red
  rectangle) is suspended in a dense granular system. The system is
  vertically vibrated with frequency $f$ and amplitude $A$.}
\label{fig_setup}
\end{figure}

Rheological properties are typically investigated by monitoring the
dynamics of a driven probe immersed in the medium. A
measure of the effective viscosity can be obtained by applying a
constant force (or torque) to the probe and measuring its stationary
angular velocity, as in the setup shown in
Fig.~\ref{fig_setup}.  The dynamics of the probe is then studied
as a function of the relevant parameters such as vibration frequency,
amplitude or velocity, mechanical properties of the materials, or
density and pressure.  One of the main aims is to pinpoint the key
quantities responsible for the different observed phenomena, such as
friction reduction and jammed-unjammed transitions in the medium.

As discussed in the previous Sections, the relevance of the study of
this kind of systems is twofold: On the one hand, one is interested in
understanding how the rheological properties of the granular medium
are modified in the presence of different kinds of perturbation; on
the other hand, the response to external stimuli is relevant to
clarify the microscopic mechanisms responsible for sliding frictional
properties or for gouge failure mechanisms (see
Sec.~\ref{model_fault}).  Concerning the first issue, the effect of flow rate and
vibration amplitude on the rheological properties of glass beads has
been recently investigated in several works, showing non-monotonic flow
curves~\cite{DWDDH11,gnoli2} and critical behaviour~\cite{WDH16}.  The effect
of mechanical fluctuations on the probe has been studied
in~\cite{PADCC15}, where a generic rheological model is proposed.
Other recent works addressed the interesting issue of non-local
rheology (see e.g.~\cite{BITCCA15}), discussing different choices for
the fluidity parameter, or the sound waves propagation (see
e.g.~\cite{santibanez2016}).

Regarding the study of the effect of mechanical vibrations on
frictional sliding, a common model system is represented by a particle
layer confined between two substrates in relative
motion~\cite{persson}, the granular medium acting as an intermediate
lubricant layer in this case.  The phenomenon of friction reduction in
this kind of systems was discovered by Capozza et
al.~\cite{CVVZ09,CVVZ12} in numerical simulations of repulsive
particles confined between a top horizontally driven plate and a
bottom vertically vibrated substrate.  In particular, focusing on the
response of the driven plate upon varying the vertical vibration
frequency, they observed suppression of friction in a well-defined
range of frequencies.  In the studied model, the vertical coordinate
$Z$ of the flat bottom substrate follows the law
\begin{equation}
Z(t)=A \sin(2\pi f t),
\label{zmax}
\end{equation}
while the
top plate is driven through a spring of elastic constant $K$ moving at
constant velocity $V$, similarly to the model described in
Sec.~\ref{model_fault}.  The friction coefficient is defined as
$\mu=F_L/F_N$, where $F_L=K[X(t)-V t]$, with $X(t)$ the horizontal
coordinate of the top plate and $F_N$ a normal external force
acting on it. The average friction coefficient $\langle \mu\rangle$
shows a non-monotonic behaviour as a function of $f$, with a marked decrease
in the interval $f\in[f_1,f_2]$. Friction reduction in a well-defined
range of frequencies shares strong similarities with the phenomenon
observed in the spring block model C discussed above. The first
transition frequency $f_1$ is related to the detachment condition and
can be predicted imposing that the inertial force $F_{in}=M\ddot{Z}$,
where $M=M_p+M_{top}$ with $M_p$ the mass of the particle layer and
$M_{top}$ the mass of the top plate, overcomes the sum of the normal
load $F_N$ and the damping force $F_{damp}=M_p\eta \dot{Z}$, where
$\eta$ is the damping coefficient accounting for viscous dissipation.
This condition gives~\cite{CVVZ09}
\begin{equation}
f_1=\frac{\eta}{2}\left(\frac{M_p}{M}+\sqrt{\frac{M_p^2}{M^2}+4\frac{F_N}{
MA\eta^2}} \right).
\label{zap1}
\end{equation}
The viscosity recovery frequency $f_2$ is obtained requiring that the
detachment time from the bottom plate equals the period of the
external oscillation and is related to the condition of maximum
momentum transfer from the vibrating plate to the confined
particles. This leads to
\begin{equation}
f_2=\sqrt{2\pi\frac{F_N}{MA}}.
\label{zap2}
\end{equation}
These theoretical predictions are in very good agreement
with numerical simulations and indicate that friction suppression is
related to the reduction of effective interface contacts in the system
due to the external vibrations. 
Finally, let us mention that recent studies~\cite{LGV15,Lasta2} have highlighted
the role played by the velocity of the imposed mechanical vibrations
on the frictional properties of sheared granular media.

%\subsection{Controlled viscosity}
\label{exp}

\begin{figure}[!htb]
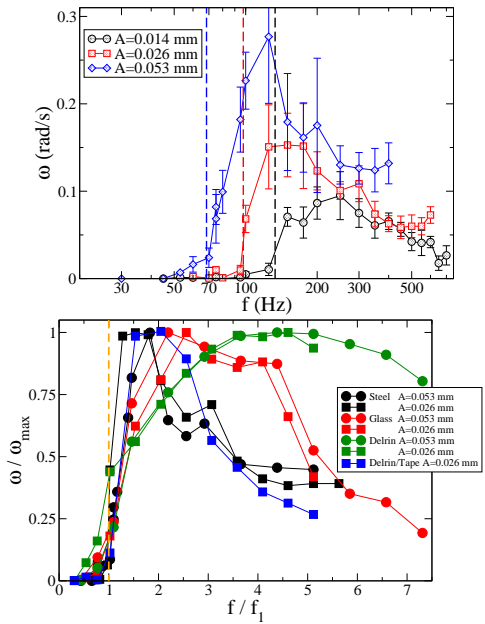

\includegraphics[width=2.2in,clip=true]{omega_f.eps}
\includegraphics[width=2.5in,clip=true]{omega_f1_nolog.eps}
\caption{(Top panel) Average angular velocities $\omega$ as a
  function of $f$ for steel beads. The vertical dashed lines represent
  the theoretical predictions for $f_1$,
  see Eq.~(\ref{f1}). (Bottom panel) Rescaled angular velocity, as a
  function of $f/f_1$, for different materials. Data points have
  standard deviation error $\sim 15\%$. }
\label{fig_exp1}
\end{figure}

The friction reduction phenomenon occurring in a range of vibration
frequencies in vertically shaken granular systems has been recently
observed in experiments~\cite{prl2018}. The setup is shown in
Fig.~\ref{fig_setup}: The granular particles (steel, glass
or delrin spheres) are confined by an aluminium plate, with packing
fraction $\sim 49\div52\%$. The system is vertically vibrated by an
electrodynamic shaker following Eq.~(\ref{zmax}), where $Z$ is the
coordinate of the shaker plate. The probe is represented by a
Plexiglas vane suspended in the medium and subject to an external
torque.  Further details on the experimental setup can be found
in~\cite{gnoli,gnoli2}. For constant applied torque, the
average angular velocity $\omega$ of the vane is proportional to the
inverse of the macroscopic viscosity of the system and therefore provides
information on its frictional properties.  The measured values of
$\omega$ are reported in Fig.~\ref{fig_exp1} (left), as a function of
$f$, for three values of the vibration amplitude $A$.

In agreement with the scenario described in
Subsections~\ref{single_block} and~\ref{zap}, one observes two regimes,
at low and high $f$, respectively. The first is characterized by high viscosity of the medium
(corresponding to small values of $\omega$), whereas in the second fluidized regime,
for intermediate values of $f$, $\omega$ reaches its maximum value
$\omega_{max}$, corresponding to viscosity reduction, see
Fig.~\ref{fig_exp1} (left). The frequency threshold $f_1$ of the first
transition from the solid (high frictional) state to the fluid (low
frictional) state is very well estimated by the theory presented in
Section~\ref{zap}, leading to Eq.~(\ref{zap1}).  In the experimental
setup, the largest force provided by the shaker is
$F_{in}=M\ddot{Z}_{max}$, with $\ddot{Z}_{max}=A (2\pi f)^2$, and
$F_N=Mg$, with $M$ the total mass of the system (granular particles
and aluminium plate).  Thus, from Eq.~(\ref{zap1}), one has
\begin{equation}
2\pi f_1 = \sqrt{g/A},
\label{f1}
\end{equation}
that is in very good agreement with the experimental results, see
dashed lines in Fig.~\ref{fig_exp1} (left).  The generality of the
underlying mechanism, related to the detachment condition, is
demonstrated by experiments with different granular materials,
reported in Fig.~\ref{fig_exp1} (right). It is interesting to note that this
fluidization phenomenon is different from the acoustic fluidization
discussed in Section~\ref{acoustic_flu}.

Increasing the vibration frequency one observes viscosity recovery
for, say, $f\gtrsim f_2$, as shown in Fig.~\ref{fig_exp1} (left). This
phenomenon is analogous to the one observed in the numerical simulations
described in Subsec.~\ref{zap}.  However, the main underlying physical
mechanism is different.  
Indeed, the viscosity recovery frequency shows a strong dependence on
the material properties, see Fig.~\ref{fig_exp1} (right).  The
physical mechanism responsible for the phenomenon relies on a balance
between dissipative and inertial forces and on the dissipation rate
in the system, analogously to the case of spring-block model discussed
in Section~\ref{single_block}. The dissipation rate is affected by the
dissipative forces characterizing both the grain-grain and the
grain-interface interactions.
As shown in Fig.~\ref{fig_exp1} (right), the recovery frequency is
significantly reduced in the case where the bottom plate is covered
with a thick layer of rubber tape (compare blue squares to green
squares), namely $f_2$ decreases upon increasing the dissipation in
the system.

\begin{figure}[!tb]
\includegraphics[width=2.5in,clip=true]{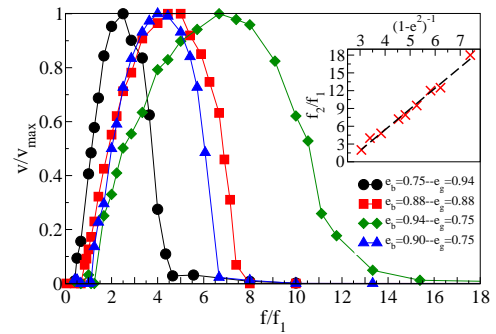}
\caption{Rescaled velocity of the probe as a function of $f/f_1$ in
  numerical simulations. Inset: The recovery frequency $f_2$ as a
  function of the inverse of the dissipation factor $1-e^2$ for
  systems with $e_g=e_b$.}
\label{fig_exp3}
\end{figure}

The role of the dissipation rate has been investigated in more detail
in numerical simulations~\cite{prl2018}, considering a geometry where
the granular medium is enclosed between two plates, the bottom one
oscillating according to Eq.~(\ref{zmax}), and confined by the
gravitational force.  As a probe, a rigid cross-shaped subset of
grains, subject to a constant force $F$ along the horizontal
direction, is used.  The velocity $v$ of the probe in the force
direction for different $f$ and $A$ shows the fluidization transition
at $f_1$ and the viscosity recovery at higher frequencies (see
Fig.~\ref{fig_exp3}).  The viscoelastic properties of the system are
explored by changing the restitution coefficient of each grain
$e$~\cite{silbert}. In particular, two different restitution
coefficients can be introduced to model the experimental situation:
$e_{g}$ for grain-grain collisions, and $e_{b}$ for collisions between
grain and bottom plate. The behaviour of $v/v_{max}$ as a function of
$f/f_1$ for different values of $e_{g}$ and $e_{b}$ is shown in
Fig.~\ref{fig_exp3}: $f_1$ is independent of $e_b$ and
$e_g$, in agreement with Eq.~(\ref{f1}), whereas the recovery
frequency $f_2$ depends on the dissipation properties.

Insights in the physical mechanisms at the origin of viscosity
recovery in this system can be obtained by the following reasoning.
Viscosity recovery occurs when the rate of energy dissipation becomes
larger than the rate of energy provided to the granular medium by the
bottom plate oscillations. Indicating with $E_{in}$ the average energy
input to the grains, and considering for simplicity a unique value of
$e$, the rate of energy dissipation is given by $\dot{E}_{dis} \sim
E_{in}(1-e^2)f_c$, where $f_c$ is the collision frequency. It is
reasonable to assume that $f_c$ is proportional to the vibration
frequency $f$, yielding $\dot{E}_{dis} \sim (1-e^2)f E_{in}$. On the
other hand, at sufficiently large frequency, the rate of energy input
$\dot{E}_{in}$ is controlled by the oscillation amplitude of the
bottom plate and becomes independent of the vibration frequency
$f$. 
%% This can be analytically shown in a simple two-block model,
%% similar to the one described in section~\ref{single_block}: a mass
%% $m_1$ with vertical coordinate $z_1$ is constrained to oscillate
%% periodically, namely $z_1=A \sin(2\pi f t)$, and is connected through
%% a spring with elastic constant $k$ to a second mass $m_2$ with
%% position $z_2$. The second mass is subject to gravity, and elastically
%% interacts with the first mass only when the distance $\Delta
%% z=z_1-z_2$ is smaller than a given length $l$.  In the absence of any
%% form of dissipation, one can show that the characteristic frequency of
%% the model is $2\pi f_0=\sqrt{k/m_2}$. For $f \gg f_0$, $m_2$ exhibits
%% oscillations of a certain amplitude and frequency $f' \simeq f_0$,
%% both independent of $f$.  
As a consequence, also $E_{in}$ and $\dot{E}_{in}$ are $f$-independent
and the condition $\dot{E}_{dis} \sim\dot{E}_{in}$ gives a recovery
frequency $f_2 \sim (1-e^2)^{-1}$.  This dependence of $f_2$ on $e$ is
confirmed by numerical simulations in the case $e=e_g=e_b$, see inset
of Fig.~\ref{fig_exp3}, showing the behaviour $f_2/f_1 \sim
(1-e^2)^{-1}$.  This analysis confirms that the second transition,
leading to viscous friction recovery, relies on the dissipation
mechanisms in the medium and between the medium and the bottom plate,
explaining the observed dependence on the materials.

\section{Acoustic Fluidization and Dynamical weakening}
\label{acoustic_flu}

In this Section, the AF mechanism is considered as a possible explanation for
the weakness of seismic faults. Indeed, many seismic faults
exhibit a resistance to shear stress which is astonishingly lower than
the one measured in laboratory experiments on both intact or ruptured
rock specimens. AF represents an explanation for the observed weakness
and can be considered an alternative to other interpretations, mostly
based on the presence of water inside the fault zone.  These
interpretations, however, are in disagreement with measurements of
porosity changes in sheared rocks \cite{Mel96}. On the other hand, the
AF scenario does not invoke the presence of water or fluids but
assumes the existence of an in-cohesive region inside the fault. If
strong elastic (acoustic) waves perpendicular to the fault plane are
activated, the normal stress is reduced and the in-cohesive region can
slide under much smaller shear stress than required in the absence of
vibrations.  This process is based on the heterogeneous and noisy
nature of the fracture propagation where the small wavelength
component of the seismic radiation is scattered by small-scale
heterogeneities inside the fault and eventually generates
perpendicular oscillations. Melosh proposes a feedback mechanism:
Vibrations promote the failure of a limited region of the fault which
slips releasing its internally stored elastic energy. However, a
fraction ${\cal E}$ of this energy becomes available to activate
additional vibrations causing further failures. The
self-sustainability of this process is based on the energetic
balance between ${\cal E}$ and the fraction ${\cal E}_{diss}$ of
energy scattered outside the fault or converted in heat. Using
experimental values for ${\cal E}_{diss}$ and ${\cal E}$, Melosh has
shown that acoustic standing waves of amplitude comparable to the
confining pressure can exist in real seismic faults. Nevertheless,
some initial acoustic energy must exists to trigger this cascading
process. Melosh invokes stress-drop events in a limited area of
the fault, generating a vibration over a sufficiently large
volume, to allow acoustic energy to initially grow and 
self-sustain. The necessity of sufficiently large volumes is also proposed
by Melosh as an explanation for the absence of AF in laboratory
experiments but deeper insights on the initiation process are still missing.
Furthermore, not only laboratory experiments but also instrumental
measurements of radiated patterns of real earthquakes cannot provide a
clear proof of the AF scenario. Indeed short wavelengths responsible
for AF cannot propagate far from the fault zone and therefore can be
observed only by instruments very close (few meters) to the rupture
surface. Numerical simulations probably represent the only approach
where the AF hypothesis can be concretely tested. The remaining part
of this Section is devoted to the study of the AF scenario in
numerical simulations of the fault model introduced in Sec.~\ref{model_fault}.

\subsection{Laboratory observation of AF}
A laboratory investigation of the AF hypothesis has been conducted in  
ref. \cite{XHM13}. In their experimental settings, acoustic excitation via compressive stress pulses, generated by high speed impact, are applied to 
a granular sample used to simulate fault gouge. Xia et al. have studied the stress-strain curve of the sample for different thickness of the granular sample as well as for different choices of the striker impact velocity $V_{im}$. 
For a fixed thickness the experiment shows a dramatic change in the rheological properties at a characteristic impact velocity $V_c$ such that the granular sample exhibits a solid-like behaviour for $V_{im}>V_c$ and a fluid-like behaviour for $V_{im}<V_c$. This result is consistent with the AF scenario since 
larger values of $V_{im}$ cause an increase of the confinement stress.
As a consequence, for  $V_{im}>V_c$ the internal pressure produced by acoustic oscillation of grains is no longer able to counterbalance the confining pressure and a solid-like behaviour is observed. This interpretation is supported by the decrease of $V_c$ as function of the granular thickness, consistently with the AF scenario, as well as its decrease when pre-stress granular samples are considered.   
An indirect evidence of AF in laboratory investigation is the clear reduction of the granular layer thickness produced by acoustic vibration and documented in ref.\cite{VEBBJ12}. This reduction, indeed, has been attributed to vibrations generated inside the sample which lead to the auto-acoustic compaction according to a mechanism similar to what predicted by the AF hypothesis.

\subsection{The AF frequency in the numerical model of seismic faults}
\label{omegaAF}

In this Section it is explored the possibility that vibrations
perpendicular to the fault plane can activate inside the fault and are
able to self-sustain. The most natural hypothesis is that these
vibrations correspond to standing waves that travel vertically
between the two confining plates. If these waves propagate with 
velocity $v_a$ their frequency is given by
$\omega_{AF}=\pi v_a/W$, where $W$ is the width of the layer. In order to provide an expression
for $\omega_{AF}$ in terms of the model parameters, we consider that the
propagation velocity is $v_a=\sqrt{M/\rho}$, where $M$ is the
$P$-wave modulus and $\rho$ the system density.  The evaluation of
$M$ in confined granular media is very complicated and, indeed,
experimental and numerical studies \cite{Makse04,ROLJJ15} indicate
that it increases for increasing confining pressure.  However,
since the shear modulus of a granular packing is negligible, compared
to the bulk modulus, in first approximation $M$ coincides with the
bulk modulus of a single grain. In the mechanical model each grain
under normal compression can be considered to deform as a cube. More precisely, a
compressional stress $\sigma_{ii}$ applied in the $i$-th direction, on
the two faces perpendicular to the $i$-th direction, produces a
deformation $2 \delta x_i$, along the $i$-th direction, with $\delta
x_i$ given by $ k_n \delta x_i= \sigma_{ii} d^2 .$ For $\delta x_i \ll
d$, one has $ \frac{\delta V}{V}=-2 \sum_{i=1}^3 \frac{\delta
  x_i}{d}=-\frac{2d}{k_n} \sum_{i=1}^3 \sigma_{ii}= -\frac{6d}{k_n}
P $, where $P$ is the applied pressure and $V$ the volume. The $P$-wave modulus then
is equal to $M=k_n/(6d)$, leading to the AF frequency
\begin{equation}
\omega_{AF}= \frac{\pi}{W} \sqrt{\frac{k_n}{6\rho d}}.
\label{oaf}
\end{equation}

\subsection{Identification of AF in the numerical model}

\subsubsection{The response to an external perturbation}
\label{secPi}

\begin{figure}[t!]
\includegraphics*[width=2.5in,clip=true]{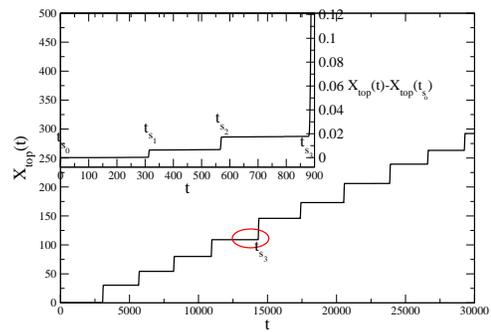}
\caption{Time evolution of the top plate position. The main panel shows the evolution
in a large time interval,
while the inset focuses on a shorter interval (red circled in the main 
panel) after the time $t_{s_0}=0$.
In this shorter interval three small slips are detected, at time $t_{s_0}$, 
$t_{s_1}=321$ and $t_{s_2}=576$,
and a large one at time $t_{s_3}=890$.}
\label{figE1}
\end{figure}

In this Section the response of the system to an external perturbation is
discussed as function of the perturbation frequency \cite{GSdALP15}. A granular fault of width $W=10d$, confined by $P_0$,
with normal spring constants $k_n=2 10^3 P_0d$ is subject to an external perturbation at frequency 
$\omega^*=1.4\pi$ which, according to Eq.(\ref{oaf}) corresponds to $\omega_{AF}$in our model.
More precisely, at a certain time $t$ the external drive is stopped, i.e. $V$ is set to zero, and for a temporal period $\tau$  
the confining pressure is changed of a quantity 
$P_{-}(t_p,t)= - \frac{\alpha
P_{0}}{2}\left[1-\sin(\frac{\pi}{2}+\omega(t_p-t)) \right ]$, with 
$t_p\in[t,t+\tau]$ and $\omega=\omega^*$. The duration of each perturbation is fixed to $\tau=10$,
which for varying frequencies leads to $n=\tau\omega/2\pi\in [1, 10^3]$ pulses. 
The reduction of the confining pressure is expected to promote a displacement of the top plate
$\Delta x(t)=x_\alpha(t+\tau)-x_0(t+\tau)$,
where $x_{\alpha}(t+\tau)$ is the position of
the top plate after
the perturbation has been applied, and $x_0(t+\tau) \simeq x_0(t)$ is the
unperturbed top plate position. The response to the external perturbation is 
then quantified by the function $\Pi_-(t,\omega^*) = \Delta x(t)/\alpha P_0$ 
which, for sufficiently small values of  $\alpha$ ($\alpha < 0.05$), becomes  
$\alpha$-independent. The evolution of $\Pi_-(t)$ is monitored in the temporal
window $[t_{s_0}:t_{s_3}]$ (Fig.~\ref{figE1})
which presents three slips of size $\Delta x_{\rm slip}(t_{s_i})<0.1$, at times 
$t_{s_0}=0, t_{s_1}$ and $t_{s_2}$, followed by a large slip $\Delta x_{\rm 
slip}(t_{s_3})=20.3$, at time $t_{s_3}$.

\begin{figure}[t!]
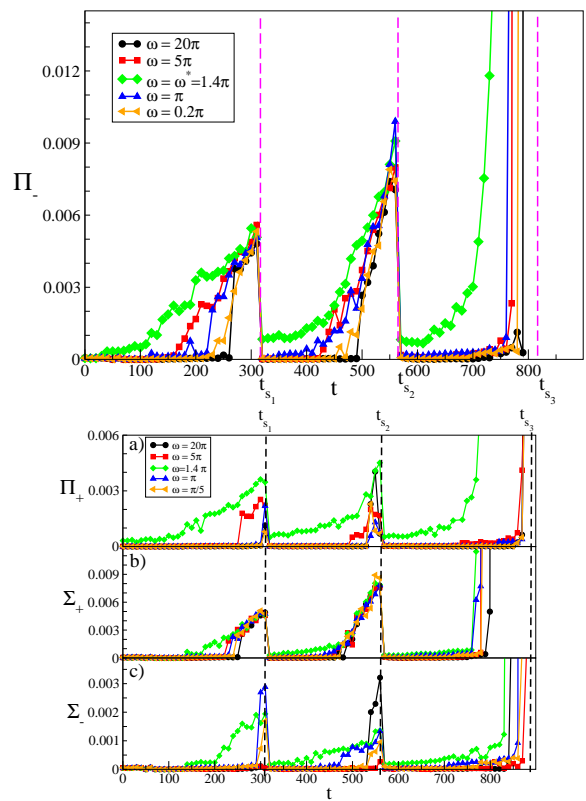

\includegraphics*[width=3in,clip=true]{figE2.eps}
\includegraphics*[width=2.5in,clip=true]{figE3.eps}
\caption{(Top panel) Time dependence of the response function $\Pi_{-}$ to the 
perturbation $P_-$ which reduces the confining pressure. Different colours 
correspond to different  frequencies. 
(Bottom panel) The response to perturbations
increasing the confining pressure (a), increasing the shear stress (b),
or decreasing the shear stress (c).}
\label{figE2}
\end{figure}

The quantity $\Pi_-(t,\omega^*)$ (left panel Fig.~\ref{figE2}) increases as $t$ 
approaches the slip occurrence time. More precisely the external 
perturbation induces a displacement of the top-plate $\Delta x(t)$ which is 
comparable to the slip in the unperturbed evolution if $t$ is sufficiently 
close to the slip occurrence time $t_{s_i}$. As a consequence, the main effect 
of the external perturbation is to anticipate the slip occurrence and one can 
define a time advance $\Delta t_a$, by the condition $\Delta x(t_s-\Delta 
t_a) = 0.2 \Delta x_{\rm slip}(t_s)$. 
This result shows that external perturbations
can reduce the confining pressure and can weaken the fault in such a way that 
slip occurs for a smaller value of the applied shear stress similar to what observed in the numerical study \cite{FGGJMC13}. 

In the following evidence is provided that the observed weakening is promoted by 
acoustic oscillations at the frequency $\omega_{AF}$, as in the  AF hypothesis, 
by showing that
\begin{itemize}
\item{(a)} A behaviour qualitatively similar to  $\Pi_-(t,\omega^*)$ is observed if 
the external perturbation increases the confining pressure or reduces the shear 
stress;
\item{(b)} The response $\Pi_-(t,\omega)$ strongly depends on the perturbation 
frequency $\omega$ and $\Pi_-(t,\omega) \simeq 0$ if $\vert \omega 
-\omega_{AF}\vert \gg 0$;
\item{(c)} The response only weakly depends on the duration $\tau$ of the applied 
perturbation.
\end{itemize}
To address item (a) the same analysis is performed under  
an external perturbation which increases the confining pressure
\begin{equation}
P_{+}(t_p,t)= + \frac{\alpha
P_{0}}{2}\left[1-\sin(\frac{\pi}{2}+\omega(t_p-t)) \right ],
\label{p+}
\end{equation}
with $t_p\in[t,t+\tau]$ or with a perturbation which forces the shear stress
to vary by $\sigma_\pm(t_p,t)= \pm \frac{\alpha
\sigma(t)}{2}\left[1-\sin(\frac{\pi}{2}+\omega (t_p-t)) \right ]$.
As in the definition of $\Pi_-$, the response of the system in the linear regime 
is quantified by $\Pi_\pm(t) = \Delta x(t)/\alpha P_0$ and $\Sigma_\pm(t) = 
\Delta x(t)/\alpha \sigma(t)$.
Results (right panel Fig. (\ref{figE2})) show that the system response to the external 
perturbation
does not depend on the orientation of the perturbation and even perturbations 
which should stabilize the system as $P_+$ or $\sigma_-$ are able to anticipate 
the slip occurrence.

\begin{figure}[t!]
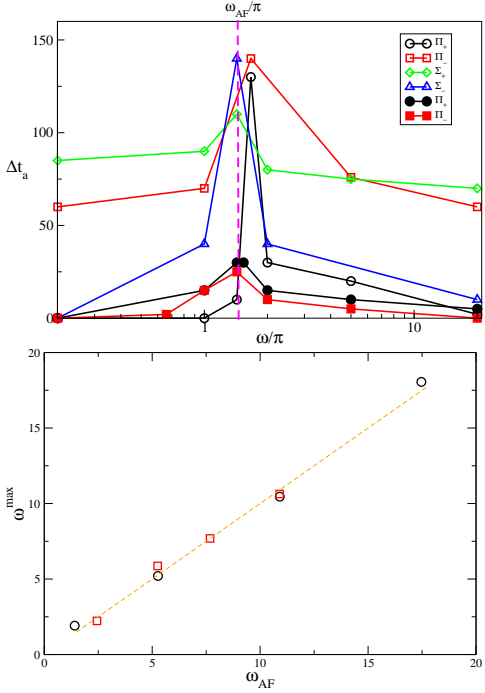

\includegraphics*[width=2.5in,clip=true]{figE4.eps}
\includegraphics*[width=2.5in,clip=true]{figE5.eps}
\caption{
(Top panel)
Frequency dependence of the advance time $\Delta t_a$
for different perturbations $P_{\pm}$ and $\sigma_{\pm}$, as in the legend,
applied for a time $\tau=10/\omega_{AF}$ (open symbols).
Filled symbols represent the response to perturbations $P_{\pm}$
applied for a time $\tau=1/\omega_{AF}$. (Bottom panel) The frequency $\omega^{max}$ 
versus $\omega_{AF}$ for different values of $W=5,7,10,20d$ (red squares)
and different values of $k_n=0.1,1,4,10 \times \overline {k_n}$ (black circles) 
with $ \overline {k_n}= 2\cdot 10^{3}\, k_d$. 
The orange dashed line corresponds to  $\omega_{max}=\omega_{AF}$.}
\label{figE4}
\end{figure}

Concerning item (b) in Fig. (\ref{figE2}) 
$\Pi_\pm(t,\omega)$ and $\Sigma_\pm(t,\omega)$ are plotted for different values of $\omega$. 
Results clearly show that the larger the difference between $\omega$ and 
$\omega^*$ the weaker is the response. This effect can be quantified by the 
dependence of $\omega$ on the advance time $\Delta t_a$ (left panel Fig.\ref{figE4}). For 
all kinds of perturbation $\Delta t_a$ presents a peak at $\omega=\omega^*$. In 
particular for $\Pi_+$ and $\Sigma_-$, $\Delta t_a$ approaches a Dirac-delta 
function indicating that the system is weakened only by perturbation at the 
frequency $\omega^*=\omega_{AF}$.
Finally, to support item $(c)$ $\Pi_\pm(t,\omega)$ and 
$\Sigma_\pm(t,\omega)$ are evaluated under perturbations of different durations $\tau$. Results 
plotted in Fig.(\ref{figE4}) confirm that $\Delta t_a$ only weakly depends on 
$\tau$.

In order to verify that fault weakness is promoted by the acoustic oscillations
described in Subsec.\ref{omegaAF}, the same analysis is performed considering 
systems of different width $W$ and with different grain elastic properties 
(different $k_n$) leading to different values of $\omega_{AF}$ according to 
Eq.(\ref{oaf}).
The frequency $\omega_{max}$ which produces the maximum value of 
the advance time $\Delta t_{a}$ (right panel Fig.\ref{figE4}) is
$\omega_{max} \simeq \omega_{AF}$ for all values of $k_n$ and $W$ clearly 
supporting the AF scenario.
Finally, it is important to emphasize that item (a) indicates that AF can be triggered 
by perturbations applied along any direction. This provides a possible
explanation of triggering caused by transient seismic waves regardless
the fault orientation.

\begin{figure}[t!]
\includegraphics*[width=2.5in,clip=true]{figE6.eps}
\includegraphics*[height=2.2in,width=2.5in,clip=true]{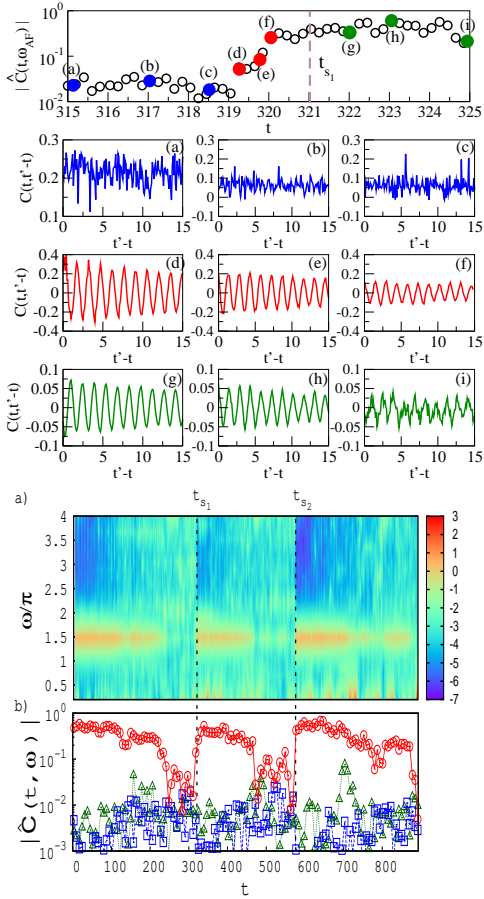}
\caption{(Top panel) Time dependence of $|\hat C(t,\omega_{AF})|$
at the characteristic frequency $\omega_{AF}$ in small temporal
windows close to slips occurring at time $t_{s_1}$.
 Panels from (a) to (i)  illustrate the temporal evolution of
$C(t,t')$  at different times identified by the same letter in the upper panel.
(Bottom panels) (a) Contour plot illustrating the time dependence of the 
logarithm of $|\hat C(t,\omega)|$. (b) Time dependence of $|\hat C(t,\omega)|$
at three different frequencies, $\omega\simeq \pi$ (squares),
$\omega=\omega_{AF}\simeq 1.5\pi$ (circles) and $\omega\simeq 2\pi$ (triangles).
The vertical lines indicate the slip occurrence time $t_{s_1}$ and $t_{s_2}$.}
\label{figE6}
\end{figure}

\subsubsection{Endogenous activation of acoustic oscillations}

In this Section, following ref.\cite{GSdALP15}, it is explored whether oscillations at the  frequency $\omega_{AF}$ 
can be  self-activated promoting slip instabilities at 
shear stress levels smaller than the one required in the absence of oscillations.

For this study, at each time $t$, a replica of the system is created
under constant external drive ($V=0$) and its spontaneous
relaxation is followed in the subsequent time interval.
The autocorrelation function of the particle velocities $\bf{v}_{i}$ is
evaluated as
$ C(t,t')= \frac{\sum_{i=1}^{N} \bf{v}_{i}(t)\cdot\bf{v}_{i}(t')}{\sum_{i=1}^{N} \bf{v}_{i}(t)\cdot \bf{v}_{i}(t)} $.
Given a slip with occurrence time $t_s$, in Fig.\ref{figE6} $C(t,t')$ is plotted
as function of $t'$ focusing on nine different values of $t$: Three times (blue 
circles, panels (a-c)) are located much before the slip time $t<t_s$, three 
times (red circles,  panels (d-f)) immediately before the slip instability $t 
\lesssim t_s$   and  three times (green circles,  panels (g-i)) after the slip 
time $t>t_s$.
The panels (d-f), corresponding to $t \lesssim t_s$, show that
 $C(t,t')$ presents a clear oscillating pattern with a typical frequency 
$\omega 
\simeq \omega_{AF}$.  Oscillations at the same frequency are also observed after the 
slip (panels (g-i)), but with an amplitude decreasing with increasing $t$. 
Conversely, before the slip (panels (a-c))  $C(t,t')$ exhibits an irregular 
pattern. The presence/absence of oscillations at $\omega=\omega_{AF}$ can be 
detected by the Fourier transform   $\hat C(t,\omega_{AF})$ of  $C(t,t')$ with 
respect to the variable $t'$. Results (upper left panel) show that   $\hat 
C(t,\omega_{AF})$ is very small in the short temporal period anticipating the 
slip. However, as soon as the slip time is approaching $t_s$, $\hat C(t,\omega_{AF})$ 
rapidly increases, reaches its maximum value in correspondence to the slip 
occurrence time and  then remains substantially constant. The evolution of  
$\hat C(t,\omega_{AF})$ on a longer temporal period is plotted in the lower right panel of Fig. 
\ref{figE6}. This figure confirms that $\hat C(t,\omega_{AF})$ presents a very 
fast increase as the slip instability is approaching. After the slip, 
conversely, it decays reaching small values at times $t$ distant from slip 
instabilities.
In the same figure $\hat C(t,\omega)$ is plotted for other values of 
$\omega$: If $\omega \ne \omega_{AF}$, $\hat C(t,\omega)$ is a 
noisy quantity fluctuating around a very small value. This figure clearly 
shows that characteristic oscillations appear only at a frequency $\omega \simeq 
\omega_{AF}$ at the onset of the slip instability. These oscillations then tend 
to disappear as the dynamics goes on. This pattern is confirmed by the 
contour map plot in the upper right panel of Fig. \ref{figE6}.

\subsection{Mechanisms producing AF in the numerical model of seismic faults}

\begin{figure}[!t]
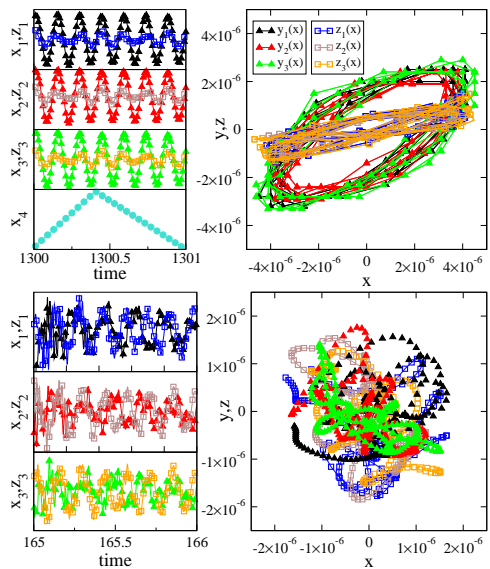

%\vspace{1cm}
\includegraphics[width=2.5in,clip=true]{figE9.eps}
\includegraphics[width=2.5in,clip=true]{figE11.eps}
\caption{(Top panels)
Time dependence of the $x$ (filled triangles) and $z$-coordinate (open squares) 
of the position of four nearest neighbour particles at the time $t_{s_1}-t =20$, 
far from the slip. The vertical scale of $x_4$ is $3000$ times larger than the 
scale of $x_1,x_2,x_3$ (first panel).  The position components  $y_1,y_2,y_3$  
and $z_1,z_2,z_3$ are plotted as a function of  $x_1,x_2,x_3$, respectively, at 
the time $t_{s_1}-t =20$. Each trajectory has been shifted to be all centred  
in $(0,0)$. The same symbols and colours as in the first panel
(second panel). (Bottom panels) The same quantities as in the top panels are plotted at the time $t_{s_1}-t =5$, i.e. at the onset of slip. }
\label{figE9}
\end{figure}

In order to identify the mechanisms responsible for the 
activation of these oscillations, the trajectories of each grain inside the stick
phase can be monitored in numerical simulations\cite{GdAPL18}.  
Because of the high
granular density, the large majority of particles is always in
contact with their neighbours forming an almost rigid structure,
i.e. the backbone. Conversely, a small fraction (less than $10\%$) of
particles, the rattlers, are located inside the cages formed by the
particles in the backbone \cite{TS10} and most of the time do not
interact with other particles, even if they contribute to stabilize the
backbone structure \cite{GdAPL17}.  In Fig. \ref{figE9} (first panel)
the $x$-position of four neighbouring particles is plotted during a short
time window far from slip: Three particles $(x_1,x_2,x_3)$ exhibit regular
oscillations along the $x$-direction.  Differently, particle $4$
is a rattler and moves along a straight line up to an abrupt change in
direction caused by a collision. As shown in the left panel of
Fig.\ref{figE9}, backbone particles exhibit an oscillating behaviour along
the $x$-direction with a characteristic frequency $\omega \simeq
\omega_{AF}$. A similar oscillating behaviour is also observed for the
$y$ and $z$ component of the particle velocity, with oscillations along
the $z$-axis much smaller in amplitude. Superimposing the centers
of each trajectory in a common point, as in Fig.\ref{figE9} (second
panel), trajectories appear to be roughly confined in a plane
and exhibit an elliptic-like shape.

\begin{figure}[!t]
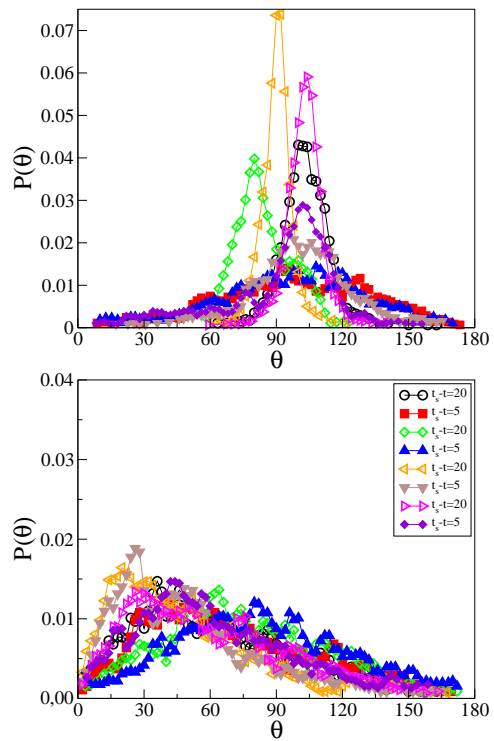

  \includegraphics[width=2.5in,clip=true]{figE10.eps}
\includegraphics[width=2.5in,clip=true]{figE12.eps}
\caption{(Top panel) The distribution of the angle $\theta$ formed by the particle 
velocities with the $z$-axis in the unperturbed evolution. We consider four slips occurring at different times $t_s$. Empty symbols are used for the distribution 
inside a temporal region sufficiently far from the slip ($t_s-t=20$) as in panels (a-c) in 
Fig.\ref{figE6}. Filled symbols are used for the distribution inside a temporal 
region at the onset of slip instability ($t_s-t=5$) as in panels (d-f) in Fig.\ref{figE6}.
(Bottom panel)  The distribution 
$P(\theta)$ when the system is perturbed by a pressure $P_+$. $P(\theta)$ is 
evaluated at the same times, identified by the same colour codes, of the left figure.}
\label{figE10}
\end{figure}

These elliptic trajectories cannot be stable during the whole stick-slip 
dynamics. Fig.\ref{figE6} indeed has shown that the correlation function 
$C(t,t')$ presents irregular behaviour with $\hat C(t,\omega) \simeq 0$ in a 
temporal window anticipating the slip instability $t_s$. This indicates that 
grain velocities decorrelate during the system evolution. The origin 
of the decorrelation is in the change of orientation of the elliptic trajectories. 
In order to prove this point,
the angle $\theta_i$ formed by the $i$-th particle velocities with  the $z$ 
axis is evaluated, together with the distribution $P(\theta)$, i.e. the number of 
particles whose velocity has orientation $\theta_i \in [\theta,\theta+\Delta \theta)$.
Fig. \ref{figE10} (left panel) shows $P(\theta)$ evaluated for different slips and at 
different times before and after $t_s$. For all slips, at large 
temporal distances from $t_s$, $P(\theta)$ is sharply peaked at $\theta \simeq 
90^o $ (open symbols), corresponding to an oscillatory motion in the $x-y$ 
plane. This distribution does not change significantly during the evolution and  
only  in proximity of the slip time it spreads towards smaller values of 
$\theta$ (filled symbols). Therefore, as $t$ approaches $t_s$  
oscillations become present also in the direction  parallel to the $z$-axis 
($\theta \simeq 0$). This is confirmed by the behaviour of the $z$-coordinate as 
function of time and as function of the $x$-coordinate (Fig.\ref{figE9}). Far from 
the slip (first panel of Fig.\ref{figE9}), the displacement in the $z$-direction 
presents oscillation at the frequency $\omega_{AF}$. As already observed 
$z-$displacements
are small compared to the $x-$displacements and the trajectory is mostly 
confined in the $x$-$y$ plane ($\theta=90^o$) (second panel of Fig.\ref{figE9}). At the onset of 
slip instability   (third and fourth panel of Fig.\ref{figE9}) 
the angle $\theta$ is no longer stable 
and $z-$ displacements of size comparable to $x-$displacements are indeed observed.
The above findings support the hypothesis of
weakening by AF. Indeed when oscillations are confined in the $x$-$y$ plane
($\theta \simeq 90^o$) they do not affect the confining pressure. Conversely, 
when $\theta \simeq 0 ^o$, oscillations can reduce the confining pressure 
promoting failure.

The overall picture is confirmed by the behaviour of $P(\theta)$ (right
panel of Fig.\ref{figE10}) when the perturbation $P_+(t,\omega)$
(Eq.\ref{p+}) is applied at the resonant frequency $\omega_{AF}$, with
$\alpha=0.05$. Differently from the unperturbed evolution when
$P(\theta)$ dramatically changes as the slip instability is
approaching (left panel of Fig.\ref{figE10}), in the presence of the
perturbation $P(\theta)$ is substantially time independent (right
panel of Fig.\ref{figE10}). In particular the presence of vertical
oscillation $\theta \simeq 0$ are observed at all times.  Combining
this observation with the behaviour of $\Pi_+$ (Subsec.~\ref{secPi}), 
it is possible to conclude that external perturbations at the frequency $\omega
\simeq \omega_{AF}$ induce the alignment of the orientation of ellipses along the
vertical direction reducing the confining pressure and promoting the
time-advance in the slip occurrence.  Interestingly, 
close to the slip instability, only small differences are found in the
angle distribution $P(\theta)$ with or without the perturbation. This
indicates similarities in the mechanisms leading to the endogenous
activation of vertical oscillations in the unperturbed evolution and
the activation triggered by the applied perturbation.

\section{Conclusions}
Occasionally, a major earthquake can trigger additional quakes up to 1000 km from
the epicentre of the first event. How that happens is not clear because the strength of the seismic
waves decreases the farther they travel. A numerical approach proposes an explanation for this observation
in terms of seismic waves from the original earthquake that can create a lathering effect in
the grains of the gouge between tectonic plates.
The fluidization of the granular medium reduces the friction between the plates, which
increases the likelihood of slipping. The model reveals that the frequency of the seismic
waves is the only important variable. Even if the amplitude of the seismic waves is very small, 
regardless their propagation direction, they are able to trigger an earthquake if their
frequency matches the resonance frequency of the fault gouge $\omega_{AF}$.
Results of Sec.~\ref{acoustic_flu} indicate that vibrational modes at the characteristic
frequency $\omega_{AF}$ do not form at the onset of slips but are
already present inside the system at all times. Reasonably, the energy
responsible for these oscillations originates from the energy stored,
during the stick phase, through the spring which couples the system to
the external drive.  Most of this energy is released very rapidly
during the slip but a significant fraction contributes to the
activation of harmonic oscillations.  Because of the vertical
confinement, only the mode at the frequency $\omega_{AF}$
(Eq.(\ref{oaf})) can be a standing wave inside the fault.  Even if
these modes have been explained in terms of compressional waves
propagating along the $z-$direction, because of the heterogeneous
structure of the granular packing, these waves induce also
displacements along the $x-$ and $y-$ directions. Far from the slip,
the confinement along the $z-$direction and periodic boundary
conditions along $x$ and $y$, lead to $x-$ and $y-$ displacements
larger than $z$-displacements. Conversely, as the system evolves there
exists a finite probability that ellipses rotate activating oscillations
along the $z-$direction. A possible mechanism
responsible for the ellipse rotation can be identified in the
collisions of rattlers with backbone particles. Indeed, these
collisions can be sufficiently energetic to destabilize oscillations
originally confined in the $x-y$ plane. Within this scenario, it is
impossible to forecast in advance the occurrence time $t_s$ of the
next slip. Indeed, rattlers follow chaotic trajectories and the
occurrence time of sufficiently energetic collisions appear to be
non-predictable.

Beside the relevance for seismic occurrence, the fluidization of a granular medium
under periodic perturbations is of extreme relevance in a variety of fields,
for many phenomena, from avalanche dynamics to the manufacturing process in
material, food, and pharmaceutical industries. Both experimentally and
numerically, different frictional regimes have been observed in the system,
from very large viscosity at low vibration frequencies to
fluidized states (corresponding to viscosity reduction) at
intermediate $f$, with a viscosity recovery at higher values of
$f$. The first transition to the fluidized state is well
characterized by the detachment condition of the granular medium from the confining plate
 and is independent of the material properties. The second transition,
leading to viscosity recovery, is related to
dissipation mechanisms in the medium and between the
medium and the bottom plate and therefore shows a strong
dependence on materials. These observations suggest the possibility
to control the viscous properties of confined granular media
by tuning the shaking frequency in the system, with
important practical application in several fields, from
tribology to geophysics and the material industry.

\vskip2cm

\enlargethispage{20pt}

%\ethics{Insert ethics statement here if applicable.}

%\dataccess{Insert details of how to access any supporting data here.}

%% \aucontribute{
%% %For manuscripts with two or more authors, insert details of the 
%% %authors contributions here. This should take the form: 
%% LdA, EL, MPC, AS conceived of and designed the                                   
%% study and drafted the manuscript.
%% All authors read and approved the 
%% manuscript.}

%% \competing{The author(s) declare that they have no competing 
%% interests}

%% \funding{MPC acknowledge support from the Singapore Ministry of Education 
%% through the
%% Academic Research Fund (Tier 2) MOE2017-T2-1-066 (S) and from the National 
%% Research Foundation Singapore.}

%% \ack{We thank A. Puglisi and A. Gnoli for an experimental collaboration on the subject discussed in Sec.~\ref{pert_freq}.}

%% \disclaimer{Insert disclaimer text here if applicable.}

%%%%%%%%%% Insert bibliography here %%%%%%%%%%%%%%

\bibliographystyle{unsrt}
\bibliography{phil_biblio}

\end{document}